\documentclass[a4paper,twocolumn,11pt,accepted=2020-01-10]{quantumarticle}
\pdfoutput=1
\usepackage[utf8]{inputenc}
\usepackage[english]{babel}
\usepackage[T1]{fontenc}
\usepackage{amsmath}
\usepackage{hyperref}

\usepackage{tikz}
\usepackage{lipsum}

\usepackage{wrapfig}
\usepackage{float}
\usepackage{amssymb}
\usepackage{amsfonts}
\usepackage{amsmath}
\usepackage{cases}
\usepackage{stmaryrd}
\usepackage{tipa}
\usepackage{subfigure}
\usepackage{bm}
\usepackage{color}
\usepackage{xcolor}
\usepackage{CJK}
\usepackage{subfigure}
\usepackage{physics}
\usepackage{array}
\usepackage{multirow}
\usepackage{nopageno}
\usepackage{mathtools}
\usepackage{enumerate}
\usepackage{hyperref}

\begin{document}

\title{Quantized Three-Ion-Channel Neuron Model for Neural Action Potentials}

\author{Tasio Gonzalez-Raya}
\affiliation{Department of Physical Chemistry, University of the Basque Country UPV/EHU, Apartado 644, 48080 Bilbao, Spain}

\author{Enrique Solano}
\affiliation{Department of Physical Chemistry, University of the Basque Country UPV/EHU, Apartado 644, 48080 Bilbao, Spain}
\affiliation{International Center of Quantum Artificial Intelligence for Science and Technology (QuArtist) \\ and Physics Department, Shanghai University, 200444 Shanghai, China}
\affiliation{IKERBASQUE, Basque Foundation for Science, Maria Diaz de Haro 3, 48013 Bilbao, Spain}

\author{Mikel Sanz}
\email{mikel.sanz@ehu.eus}
\affiliation{Department of Physical Chemistry, University of the Basque Country UPV/EHU, Apartado 644, 48080 Bilbao, Spain}
\maketitle

\begin{abstract}
The Hodgkin-Huxley model describes the conduction of the nervous impulse through the axon, whose membrane's electric response can be described employing multiple connected electric circuits containing capacitors, voltage sources, and conductances. These conductances depend on previous depolarizing membrane voltages, which can be identified with a memory resistive element called memristor. Inspired by the recent quantization of the memristor, a simplified Hodgkin-Huxley model including a single ion channel has been studied in the quantum regime. Here, we study the quantization of the complete Hodgkin-Huxley model, accounting for all three ion channels, and introduce a quantum source, together with an output waveguide as the connection to a subsequent neuron. Our system consists of two memristors and one resistor, describing potassium, sodium, and chloride ion channel conductances, respectively, and a capacitor to account for the axon's membrane capacitance. We study the behavior of both ion channel conductivities and the circuit voltage, and we compare the results with those of the single channel, for a given quantum state of the source. It is remarkable that, in opposition to the single-channel model, we are able to reproduce the voltage spike in an adiabatic regime. Arguing that the circuit voltage is a quantum variable, we find a purely quantum-mechanical contribution in the system voltage's second moment. This work represents a complete study of the Hodgkin-Huxley model in the quantum regime, establishing a recipe for constructing quantum neuron networks with quantum state inputs. This paves the way for advances in hardware-based neuromorphic quantum computing, as well as quantum machine learning, which might be more efficient resource-wise. 
\end{abstract}

\section{Introduction}
An important part of the comprehension of human beings goes through unveiling how the brain functions. The field of neuroscience, and particularly neurophysiology, was born with this intent. Along the way, the combination of neurosciences with interdisciplinary research areas, such as biophysics or bioinformatics, results from an attempt to use these advances to improve our lives. In this race, the Hodgkin-Huxley model represents a milestone. Alan Lloyd Hodgkin and Andrew Fielding Huxley received the Nobel Prize in Physiology or Medicine in 1963 for describing the propagation of electric signals through the giant axon of loligo squids \cite{HHa, HHb, HHc}. This work identifies that small segments of the axon's membrane behave as connected electric circuits \cite{HHM}. Thus, the dynamics of the conduction of nervous impulses can be represented by a set of non-linear differential equations. This established a bridge between neuroscience and physics~\cite{NeuronRev}, which is supported by many works that followed until today \cite{CohResHH, CohResHHNoise, NoiseHHM, ColorNoiseHH, TemporalCohHH, TimeScalesHH, IonChannelNoiseHH, SynTimeDelayHH}.

A neuron is an electrically-excitable cell composed of dendrites, a cell body, and an axon. The first are ramifications that incorporate the incoming stimulus into the cell body, which processes it. Subsequently, the response is fired and transmitted through the axon. Propagation of electric signals is possible due to the variation of ion permeabilities in the axon's membrane due to a depolarizing impulse. This results in the change of ion concentrations inside and outside the axon membrane, represented by voltage-dependent conductances in the Hodgkin-Huxley circuit. The dependence of these elements on previous electric signals yields natural their identification with memristors \cite{Mem, HHMem}.

Nowadays, we are looking at many novel research fields that merge biosciences and quantum technologies, many of them resulting from a progress in quantum platforms achieved in the last decade. In this topic, superconducting circuits are worth mentioning. Among these fields we find quantum biology \cite{QBio, QBiobook}, quantum artificial life \cite{QAlife, QAlife_IBM}, or quantum biomimetics \cite{Biomim}. When considering the field of neurophysiology, we come across the concept of neuromorphic quantum computing~\cite{NQC}, which illustrates the attempt to enhance computational power by analyzing brain-inspired architectures with quantum features, such as entanglement. 

Multiple mathematical models have been produced recently in an attempt to condense the main features of brain neurons, as well as in an effort to extend this to the quantum realm \cite{QuanDotNeuron, QubitNeuron, MirrorNeuron, ArtificialNeuron, AAGuik, LearningNeuron}. Furthermore, synaptic and learning processes in neurons have been simulated using classical memristive devices \cite{Synap, SynapticDynamics, Neuromorphic, STDP}. A key element in the development of a quantum neuron model is the quantum memristor \cite{QMem}, and the realizability of this model lies on the proposals for constructing a quantum memristor in superconducting circuits \cite{CircuitQMem} and in integrated quantum photonics \cite{OptQMem}. A simplified version of this model has already been studied in the quantum regime \cite{QHH}. While this version accounted for the dynamics of the potassium channel solely, we consider the composite action of the potassium, sodium, and chloride channels, which represent the complete Hodgkin-Huxley model. This study is supported by previous works describing coupled memristor circuits \cite{CoupledMem, CompositeMem, TheoremMem, AnalogMem}, concerning parallely connected memristors.  

In this article, we study the quantization of the complete Hodgkin-Huxley model. This work considers potassium, sodium, and chloride ion channels, and represents a more thorough description of the conduction of nervous impulses in neurons, in a quantum regime. For this aim, we propose a setup consisting of a quantized source, the Hodgkin-Huxley circuit, and an output waveguide. We are able to quantize this circuit by introducing the quantum memristor \cite{QMem}, and we study the voltage response in the circuit, as well as in the output waveguide. This is accompanied by a simulation of the behavior of the ion-channel conductances during this process. In order to check the quantumness of the system, we compute the second moment of the voltage. At last, we reproduce the spike of the action potential in the quantum regime, with an adiabatic approximation. This work goes further into the comprehension of this quantized model, establishing the possibility of constructing quantum neuron networks able to process quantum information, as well as of the application to neuromorphic quantum architectures and quantum neural networks \cite{PercepMem}. This could find applications in the field of quantum machine learning \cite{IntroQML, QML Lloyd} without the necessity of a universal quantum computer.

\section{Theoretical Framework}
\subsection{Classical Hodgkin-Huxley model}
\begin{figure}[t]
\centering
{\includegraphics[width=0.4 \textwidth]{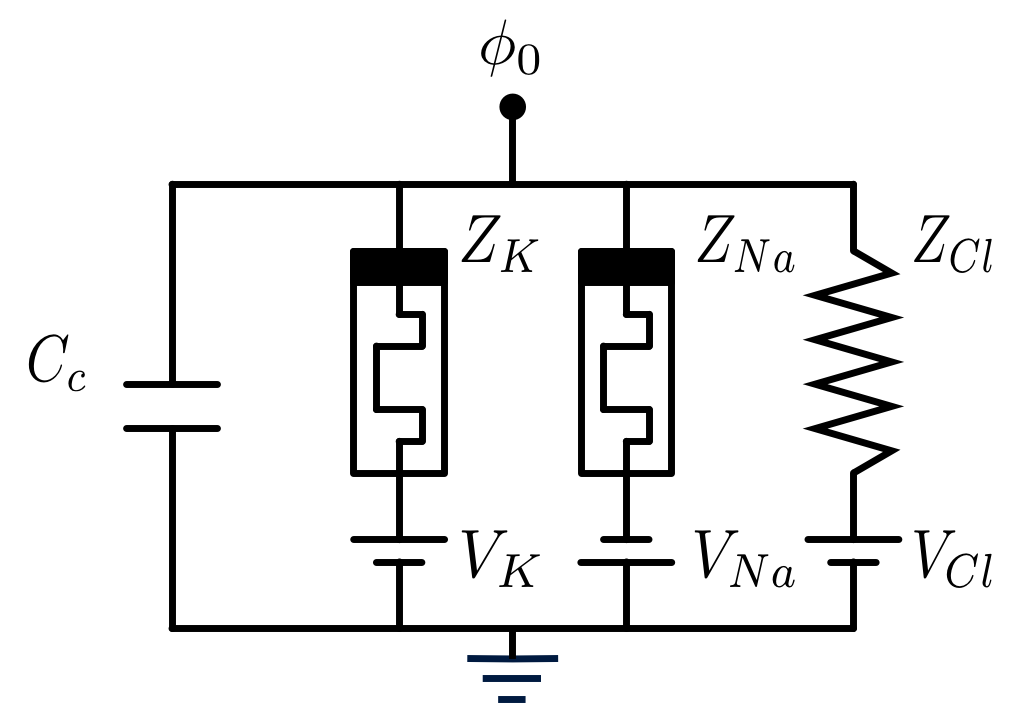}}
\caption{Complete Hodgkin-Huxley circuit featuring potassium, sodium, and chloride ion channels. The axon's membrane capacity to store charge is described by $C_{c}$. $V_{i}$ are the resting potentials representing the initial ion concentrations in the axon, and $Z_{i}$ represent the ion channel permeabilities, which are given by non-linear conductances, for $i=\text{K}, \text{Na}, \text{Cl}$. $\phi_{0}$ is the flux on the circuit, giving $\dot{\phi}_{0}$ the circuit voltage.}
\label{fig1}
\end{figure}

The reception of stimuli by a neuron causes its axon membrane to experience changes in ion permeability, subsequently leading to variations in the concentrations of ionic species in and out of the membrane. When the incoming electric signals overcome a certain threshold, the membrane is depolarized and the nervous impulse is transmitted through the axon. The ionic species that play a role in this process, potassium, sodium, and chloride, were identified by A. L. Hodgkin and A. F. Huxley, who in 1952 described the transmission of nervous impulses by means of an electric circuit. They identified each element in the axon membrane involved in this process with that of an electric circuit. The circuit which appears in the Hodgkin-Huxley model can be seen in Fig.~\ref{fig1}, and the equations describing it are
\begin{eqnarray}
\label{current} \nonumber I(t)&=&C_g\frac{d V_g}{dt}+\bar{g}_{\text{K}} n^4 (V_g-V_{\text{K}})\\
&+& \bar{g}_{\text{Na}} m^3 h (V_g-V_{\text{Na}}) +\bar{g}_{\text{Cl}}(V_g-V_{\text{Cl}}),\\
\label{mu}\frac{dn}{dt}&=&\alpha_n(V_g)(1-n)-\beta_n(V_g)n,\\
\label{m}\frac{dm}{dt}&=&\alpha_m(V_g)(1-m)-\beta_m(V_g)m,\\
\label{h}\frac{dh}{dt}&=&\alpha_h(V_g)(1-h)-\beta_h(V_g)h,
\end{eqnarray}
where $I(t)$ is the input current, $\bar{g}_i$ ($i=\text{Cl}, \text{K}, \text{Na}$) represents the maximum value of the ion-channel electrical conductance, $C_{c}$ is the axon membrane capacitance, and $V_{i}$ $(i=\text{Cl}, \text{K}, \text{Na})$ is the resting potential of the ion channel. Here, $n, m,$ and $h$ are gate-opening probabilities, which give the probability of activation for a single gate or sub-unit, requiring a given ion channel that its 4 gates be open in order for it to be activated. The chloride channel Cl, which accounts for small unperturbed flow of non-involved (mostly chloride) ions, is described by a constant conductance. On the contrary, K and Na ion channel are characterized by non-linear conductances due to \textit{n, m} and \textit{h}, which give a different weight to each channel and depend on voltage and time. The nonlinear conductances in the Hodgkin-Huxley circuit can be described by memristors. 

\subsection{Memristor}
A memristor is a dissipative circuit element whose resistance depends on the history of signals, voltages and charges, which have crossed it. This description of an ideal resistive element with memory effects can be done in Kubo's response theory \cite{Kubo}, but it was L. O. Chua who, in 1971, coined the term memristor \cite{Mem}. The equations describing this element are
\begin{eqnarray}
\label{I-V}I(t)=G(\mu(t))V(t),\\
\label{mudyn}\dot{\mu}(t)=f(\mu(t),V(t)),
\end{eqnarray}
for a voltage-controlled memristor. Given that $G(\cdot)$ and $f(\cdot)$ are continuous real functions, satisfying
\begin{enumerate} [(i)]
\item $G(\mu) \geq 0$ for all values of $\mu$.
\item For a fixed $\mu$, $f(\mu,V)$ is monotone, and $f(\mu,0)=0$.
\end{enumerate}
Property (i) implies $G(\mu)$ can be understood as a conductance, and hence Eq.~\ref{I-V} can be interpreted as a state-dependent Ohm's law. This ensures that the memristor is a passive element. Property (ii) restricts the internal variable dynamics to provide non-vanishing memory effects for all significant voltage inputs, implying that it does not have dynamics in the absence of voltage.

See that attempting to solve Eq.~\ref{mudyn} requires time integration over the past of the control signal. This means that the current response given by the voltage-controlled memristor described in Eq.~\ref{I-V} depends, through $G(\mu)$, on previous values of the control voltage, as well as on the present one. Thus, a memristor that undergoes a periodic control signal will display a hysteresis loop when plotting the response versus the control signal (current vs voltage, in this case). The slope of this curve is identified with the resistance of the device, and the area under it is associated with memory effects \cite{CircuitQMem}.

\subsection{Circuit Quantization}
\begin{figure}[t]
\centering
{\includegraphics[width=0.45 \textwidth]{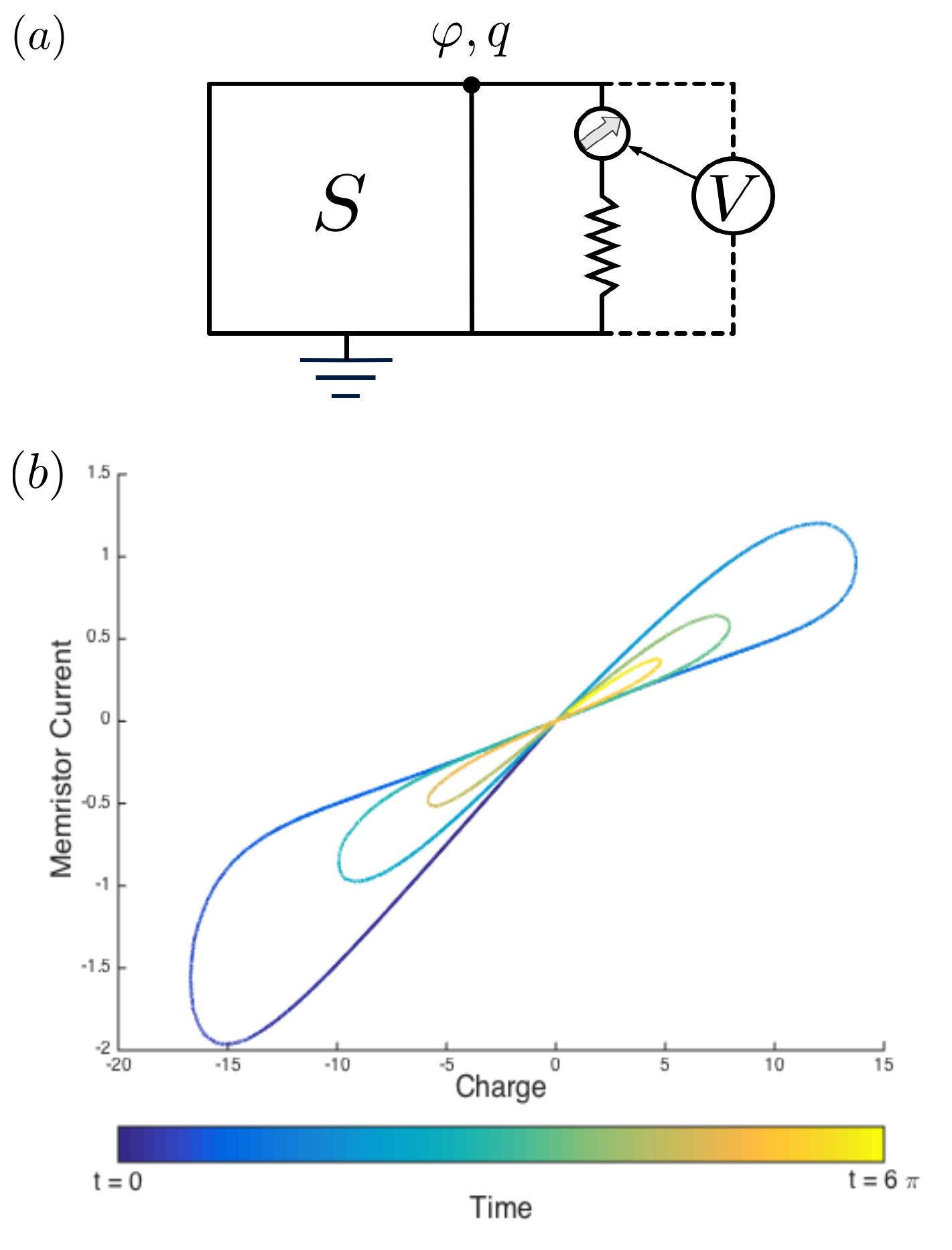}}
\caption{(a) Representation of a quantum memristor as a resistor coupled to a closed system S with a voltage-based weak-measurement scheme. (b) Memristor current vs charge for a quantum memristor coupled to a LC circuit, with a system Hamiltonian $H_{S}= \frac{q^{2}}{2C}+\frac{\varphi^{2}}{2L}$, with $C=1$F and $L=1$H. The memristor current is expressed in C/s, and the charge in C. This curve displays multiple hysteresis loops, which represent the damping of the memristor. 3 complete cycles of the memristive device are represented, from $t=0$ to $t=6\pi$.}
\label{fig2}
\end{figure}

In this section, we give a brief introduction to circuit quantization, focused on the circuit in Fig.~\ref{fig1}. This entails a description of the quantization of the memristor. We describe a resistive element with a weak-measurement scheme, used to update its resistance, as described in Ref.~\cite{QMem}. For this aim, we propose the setup in Fig.~\ref{fig2}~(a), which represents a closed system coupled to a resistor and a measurement apparatus, with a measurement-based update of the resistance depending on the voltage in the system.  

The evolution of the state of the composite system can be described by a master equation composed of a Hamiltonian part, a continuous-weak-measurement part and a classical feedback part,
\begin{equation}
d\rho=d\rho_H+d\rho_{m}+d\rho_{d}.
\end{equation}
The Hamiltonian part is given by the von Neumann equation
\begin{equation}
d\rho_{H}=-\frac{i}{\hbar}[H_{S},\rho(t)] dt.
\end{equation}
The continuous-weak-measurement part reads
\begin{eqnarray}
\nonumber d\rho_{m} &=& -\frac{\tau}{q_0^2}[q, [q, \rho(t)]]dt \\
&+&\sqrt{\frac{2\tau}{q_0^2}}(\{q, \rho(t)\}-2\langle q \rangle \rho(t))dW,
\end{eqnarray}
where $\tau$ is the projection frequency, $q_0$ is the uncertainty related to the circuit charge $q$, and $dW$ is the Wiener increment, associated to the stochasticity associated with weak measurements. The measurement strength is defined as $\kappa=\frac{\tau}{q_0^2}$, and $[\cdot, \cdot]$ denotes the commutator, and $\{\cdot, \cdot\}$ the anticommutator. The expectation value of an observable is $\langle A \rangle=\tr(\rho A)$.

The dissipation is described by Caldeira-Leggett master equation,
\begin{eqnarray}
\nonumber d\rho_{d} &=& -\frac{i\gamma(\mu)}{\hbar}[\varphi, \{q, \rho(t)\}]dt \\
&-& \frac{2C\lambda \gamma(\mu)}{\hbar}[\varphi, [\varphi, \rho(t)]]dt
\end{eqnarray}
where $\varphi$ is the flux in the circuit, $\lambda=k_{B}T/\hbar$ and $\gamma(\mu)$ is the relaxation rate. A particular solution to these equations gives the relation between memristive current and the charge shown in Fig.~\ref{fig2}~(b), for the quantum memristor coupled to a LC circuit, with a Hamiltonian of the form $H_{S}= \frac{q^{2}}{2C}+\frac{\varphi^{2}}{2L}$. In our case, the circuit is coupled to an open element (a semi-infinite transmission line), and thus there is no need to introduce the Wiener noise. 

\begin{figure*}[t]
\centering
{\includegraphics[width=1 \textwidth]{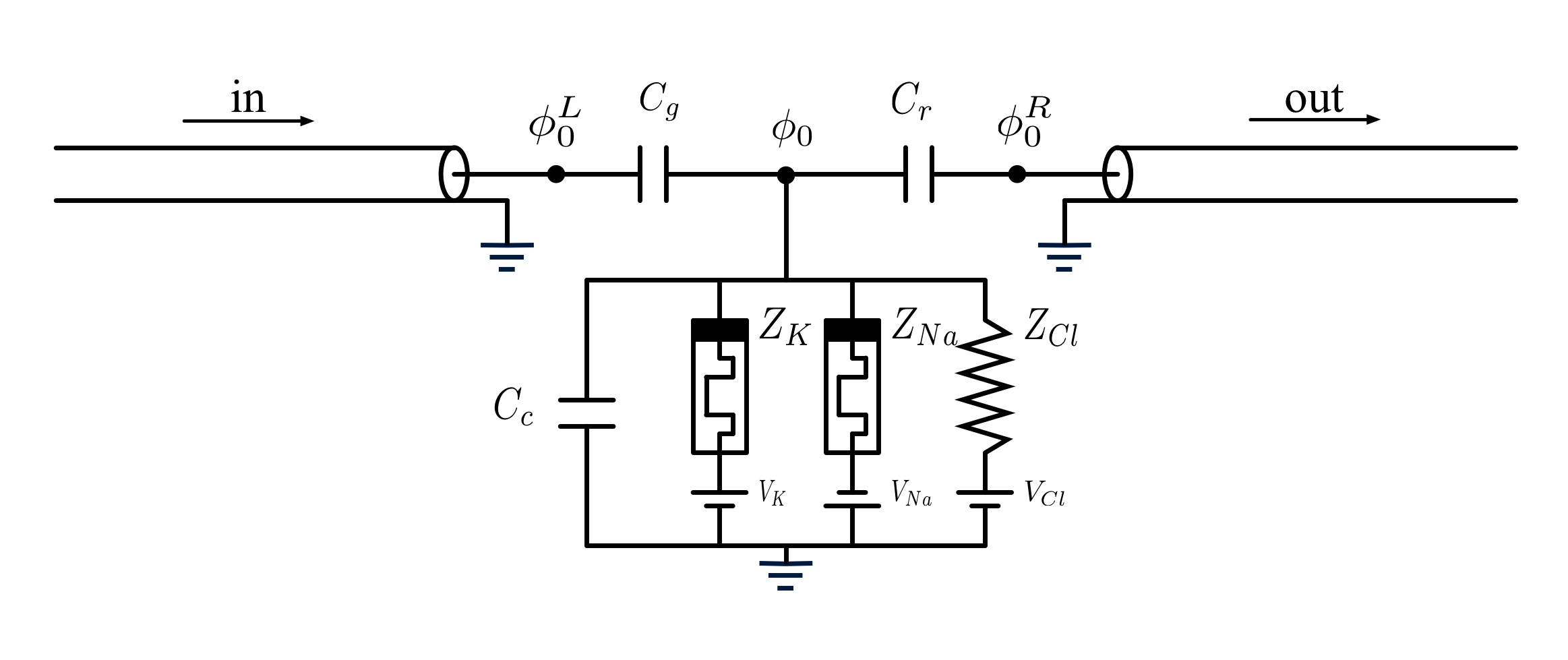}}
\caption{Complete Hodgkin-Huxley circuit with potassium, sodium, and chloride ion channels, coupled to two semi-infinite transmission lines, introducing a quantized source on the left and an outgoing waveguide on the right. $C_{g}$ is the capacitance coupling the source and the system, $C_{c}$ accounts for the axon's membrane capacitance, and $C_{r}$ coupled the system and the outgoing waveguide. $V_{K}$, $V_{Na}$, and $V_{Cl}$ are the resting potentials for the potassium, sodium, and chloride channels, and $Z_{K}$, $Z_{Na}$, and $Z_{Cl}$ are the impedances associated to the transmission lines modeling the potassium, sodium, and chloride channels, respectively. In the first two cases, those transmission lines are modeling memristors.}
\label{fig3}
\end{figure*}

Circuit quantization involves defining fluxes and charges, from which the voltage and the current can be obtained by time differentiation. To describe a circuit with linear capacitances and inductances we employ a node formulation, in which node fluxes play the role of the spatial variable. At this point, we need to find a proper description for a dissipative element, such as a memristor, in a Lagrangian formalism. This can be done by adding a dissipation function to the equations of motion of an effective Lagrangian \cite{MemLagrangian}. However, for proper canonical quantization, we need to find a Hamiltonian, and the corresponding Hamilton equations need to be time-reversible. This conflicts with the irreversibility of dissipative terms in the Lagrangian. A solution to this problem is to assume linear dissipation and treat the dissipative element in the Caldeira-Leggett model \cite{IntroQEC}. 

\begin{figure}[t]
\centering
{\includegraphics[width=0.35 \textwidth]{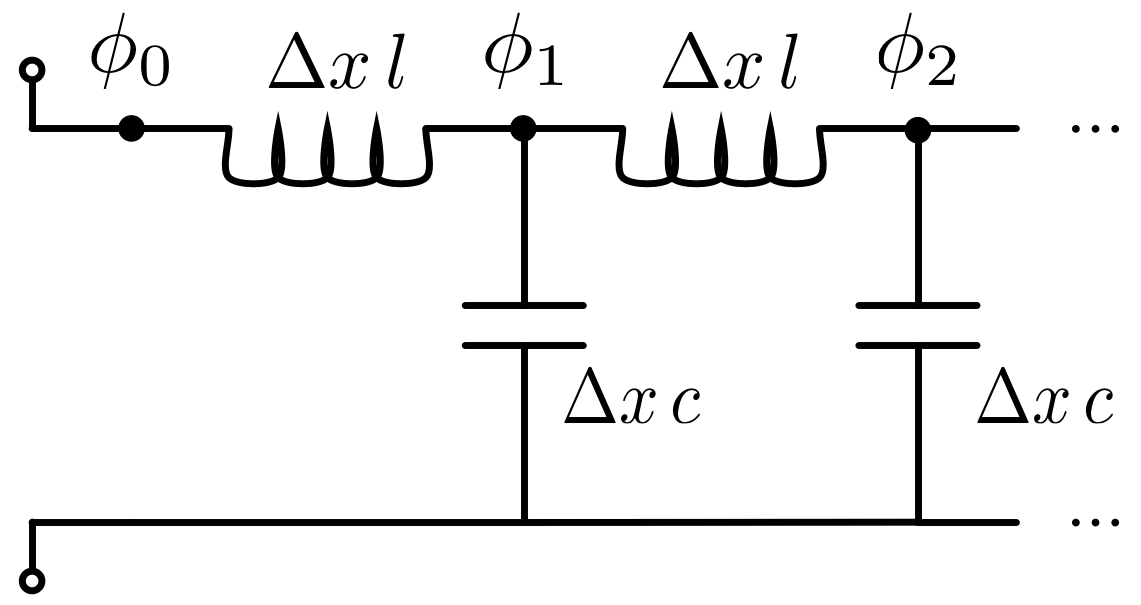}}
\caption{Transmission line modeling a linear resistor. In this setup, $\phi_{i}$ are the fluxes at each node, $\Delta x \, l$ represents the inductance, and $\Delta x \, c$ represents the capacitance, being $l$ and $c$ the density of inductance and capacitance, respectively.}
\label{fig4}
\end{figure}

In this context, we replace a linear dissipative element by an infinite set of coupled LC oscillators with a frequency-dependent impedance $Z(\omega)$, i.e. a transmission line (see Fig.~\ref{fig4}). The now infinite degrees of freedom of the resistor are identified with the intermediate node fluxes of the oscillators. This way we introduce the memristor as a linear dissipative element cast in the Caldeira-Leggett model.

\section{Complete Hodgkin-Huxley model}
The quantization of a simplified version of the Hodgkin-Huxley model, accouting solely for the dynamics of the potassium channel, has been studied in Ref.~\cite{QHH}. We complete the description given there by studying the three-ion-channel Hodgkin-Huxley model, accounting for the contribution from potassium, sodium, and chloride ion channels. We also adress the possibility of sending quantum states into the system by adding a waveguide to model the source. We even include an outgoing waveguide to account for the connection of the system onto a subsequent neuron. The aforementioned system under study is the complete Hodgkin-Huxley neuron, which includes potassium, sodium, and chloride channels modeled by two quantum memristors and a resistor, respectively. This setup is depicted in Fig.~\ref{fig3}. 

The Lagrangian describing this system is
\begin{widetext}
\begin{eqnarray}
\nonumber \mathcal{L} &=& \sum^{\infty}_{i=1}\bigg[ \frac{\Delta x c_{0}}{2}(\dot{\phi}^{L}_{i})^{2}-\frac{(\phi^{L}_{i}-\phi^{L}_{i+1})^{2}}{2\Delta x l_{0}}\bigg] -\frac{(\phi^{L}_{0} - \phi^{L}_{1})^{2}}{2\Delta x l_{0}} + \frac{C_{g}}{2}(\dot{\phi}_{0}-\dot{\phi}^{L}_{0})^{2} +\frac{C_{c}}{2}\dot{\phi}_{0}^{2} + \frac{C_{r}}{2}(\dot{\phi}^{R}_{0}-\dot{\phi}_{0})^{2} \\
\nonumber &+& \sum_{a=K,Na,Cl}\bigg\{- \frac{(\phi^{a}_{1}-\phi_{0})^{2}}{2\Delta x' l_{a}} + \sum^{\infty}_{j=1}\bigg[ \frac{\Delta x' c_{a}}{2}(\dot{\phi}^{a}_{j})^{2}-\frac{(\phi^{a}_{j+1}-\phi^{a}_{j})^{2}}{2\Delta x' l_{a}}\bigg]\bigg\} -\frac{(\phi^{R}_{1}-\phi^{R}_{0})^{2}}{2\Delta x l_{1}}  \\
&+& \sum^{\infty}_{m=1}\bigg[ \frac{\Delta x c_{1}}{2}(\dot{\phi}^{R}_{m})^{2}-\frac{(\phi^{R}_{m+1}-\phi^{R}_{m})^{2}}{2\Delta x l_{1}}\bigg]  
\end{eqnarray}
\end{widetext}
where $c_{0}$ and $l_{0}$ are the capacitance and inductance corresponding to the left transmission line, the ones corresponding to the ion channels are $c_{\text{K}}$ and $l_{\text{K}}$ for the potassium channel, $c_{\text{Na}}$ and $l_{\text{Na}}$ for the sodium channel, and $c_{\text{Cl}}$ and $l_{\text{Cl}}$ for the chloride channel, and $c_{1}$ and $l_{1}$ are the capacitance and inductance corresponding to the right transmission line. 

$\phi^{L}_{i}$ is the node flux inside the left transmission line, and $\phi^{R}_{m}$ is the node flux inside the right transmission line. On the system we have node fluxes $\phi^{a}_{j}$, for $a= \text{K}, \text{Na}, \text{Cl}$, referring to those of transmission lines modeling the potassium, sodium, and chloride channels. $\phi_{0}$ is the node flux at the point $x=0$ of the system. 

The equations of motion for this circuit are obtained as Euler-Lagrange equations, $\dv{}{t}\pdv{\mathcal{L}}{\dot{\phi}}=\pdv{\mathcal{L}}{\phi}$. For intermediate node fluxes on the transmission line, $\phi_{i}$, we find 
\begin{equation}\label{waveq}
\ddot{\phi}_{i} =  \frac{1}{l_{j} c_{j}} \pdv[2]{\phi_{i}}{x}
\end{equation}
after taking the continuum limit, $\Delta x\rightarrow 0$, and this is the wave equation for a flux field at position $x_{i}$ on the transmission line. The general solution to this equation can be written in terms of ingoing and outgoing waves, 
\begin{equation}
\phi(x,t)=\phi_\text{in}(t + x/v) + \phi_\text{out}(t - x/v),
\end{equation}
with velocity $v=1/\sqrt{l c}$. This leads to
\begin{equation}
\begin{split}
& \pdv{\phi(x,t)}{t} = \dot{\phi}_\text{in}(t+x/v) + \dot{\phi}_\text{out}(t-x/v),\\
& \pdv{\phi(x,t)}{x} = \frac{1}{v} (\dot{\phi}_\text{in}(t+x/v) - \dot{\phi}_\text{out}(t-x/v)),
\end{split}
\end{equation}
which allow us to obtain $\pdv{\phi_{0}(t)}{x} = \frac{1}{v} (2\dot{\phi}_\text{in}(t) - \dot{\phi}_{0}(t))$. Thus, we can rewrite the Euler-Lagrange equations for $\phi_{0}^{L}$, $\phi_{0}$, and $\phi_{0}^{R}$. The equation of motion for the flux field in the transmission line modeling the source is
\begin{equation}
C_{g}(\ddot{\phi}_{0}^{L}-\ddot{\phi}_{0}) = \frac{2\dot{\phi}^{L}_{\text{in}}-\dot{\phi}^{L}(x=0)}{Z_{0}},\end{equation}
and the equation for $\phi_{0}(t)$ on the circuit is
\begin{equation}
\begin{split}
&C_{g}(\ddot{\phi}_{0}-\ddot{\phi}_{0}^{L}) + C_{c} \ddot{\phi}_{0} +C_{r}(\ddot{\phi}_{0}-\ddot{\phi}_{0}^{R}) =\\
& \sum_{a=K,Na,Cl}\Big[\frac{2\dot{\phi}^{a}_{\text{in}}-\dot{\phi}^{a}(x=0)}{Z_{a}} \Big],
\end{split}
\end{equation}
where $Z_{a}$ is the impedance of the transmission line modeling the ion channel, for $a= \text{K}, \text{Na}, \text{Cl}$. The equation of motion for the flux field on the outgoing waveguide is
\begin{equation}
C_{r}(\ddot{\phi}_{0}^{R}-\ddot{\phi}_{0}) = \frac{2\dot{\phi}^{R}_{\text{in}}-\dot{\phi}^{R}(x=0)}{Z_{1}},\end{equation}
where the impedance of the final transmission line is given by $Z_{1}$. The flux field on a semi-infinite transmission line can be written in terms of ingoing and outgoing modes which satisfy canonical commutation relations. The quantization of this field has been performed, for example in Ref.~\cite{QNetwork}, for infinite electrical networks. As a starting point, we write the decomposition
\begin{eqnarray}\label{chalmers}
\nonumber \phi(x,t) &=& \sqrt{\frac{\hbar Z}{4\pi}}\int_{0}^{\infty}\frac{d\omega}{\sqrt{\omega}} ( a_{\text{in}}(\omega) e^{i(k_{\omega}x - \omega t)}+\\
&+& a_{\text{out}}(\omega) e^{-i(k_{\omega}x + \omega t)}+\text{h.c.}),
\end{eqnarray}
where $k_{\omega}=|\omega|\sqrt{l c}$ is the wave vector and $Z=\sqrt{l/c}$ is the characteristic impedance of the transmission line, associated with the resistance of the memristor. Since $a_\text{in}(\omega)$ and  $a_\text{out}(\omega)$ can be promoted to quantum operators, $\phi_{0}(t)$ is promoted to a quantum operator. Using this decomposition, we can rewrite the equations of motion for the different fields at $x=0$ in terms of ingoing and outgoing modes. Furthermore, we impose that the field in the system satisfies the following condition
\begin{equation}\label{fluxmatch}
\phi^{\text{K}}(x=0,t) = \phi^{\text{Na}}(x=0,t) = \phi^{\text{Cl}}(x=0,t),
\end{equation}
all of them equal to $\phi_{0}(t)$. Based on the results of Ref.~\cite{CoupledMem}, we identify
\begin{equation}
\frac{1}{Z} = \frac{1}{Z_{\text{K}}} + \frac{1}{Z_{\text{Na}}} + \frac{1}{Z_{\text{Cl}}},
\end{equation}
and we also define the factor related to the impedance matching,
\begin{equation}\label{theta}
\theta = \frac{\sqrt{Z_{\text{K}}Z_{\text{Na}}+Z_{\text{K}}Z_{\text{Cl}}+Z_{\text{Na}}Z_{\text{Cl}}}}{\sqrt{Z_{\text{K}}Z_{\text{Na}}}+\sqrt{Z_{\text{K}}Z_{\text{Cl}}}+\sqrt{Z_{\text{Na}}Z_{\text{Cl}}}}.
\end{equation}
This leads to the following expression:
\begin{equation*}
\begin{pmatrix}
a^{L}_{\text{out}}(\omega) \\  a_{\text{out}}(\omega) \\ a^{R}_{\text{out}}(\omega) \end{pmatrix}
= \begin{pmatrix} R_{0}(\omega) & s_{0}(\omega) & t_{0}(\omega) \\
s(\omega)  & R(\omega) &  t(\omega) \\
t_{1}(\omega) & s_{1}(\omega) & R_{1}(\omega) \end{pmatrix}
\begin{pmatrix}
a^{L}_{\text{in}}(\omega) \\ a_{\text{in}}(\omega) \\ a^{R}_{\text{in}}(\omega) \end{pmatrix}
\end{equation*}
for ingoing and outgoing modes on the transmission lines modeling the source, the output waveguide, and the ion channels in the system. The values of the reflection and transmission coefficients are presented in the Appendix~\ref{appendix}. 

We can obtain the voltage response of the system with these equations. For the sake of simplicity, we will choose the transmission lines modeling the memristors and the resistor in the system, as well as that modeling the outgoing waveguide, to be in the vacuum state. Now, the state of the source has to be chosen carefully. In Ref.~\cite{QHH}, the quantization of a simplified version of the Hodgkin-Huxley model was studied, subjected to a classical AC source, $I(t)=I_{0}\sin(\Omega t)$. For the sake of consistency, we will study the proposed system subjected to the same harmonic input. However, now the input current in the system is given by $\langle \dot{Q}^{L}_{0}(t)\rangle=C_{g}\langle\ddot{\phi}^{L}_{0}(t)-\ddot{\phi}_{0}(t)\rangle$. We find a state $|\psi\rangle$ such that $\langle \dot{Q}^{L}_{0}(t)\rangle_{\psi}=I_{0}\sin(\Omega t)$. To fulfill this claim, we propose a state of the form
\begin{equation*}
\sum_{n=0}^{\infty}\int d\omega_{1} ...\, d\omega_{n} \, f(\omega_{1}, ... \, , \omega_{n}) \, a^{\dagger}(\omega_{1}) ... \, a^{\dagger}(\omega_{n}) |0\rangle.
\end{equation*}
Choosing a single-mode state, namely
\begin{equation}
|\psi\rangle=\alpha|0\rangle+\beta\int_{0}^{\infty}d\omega f(\omega)a^{\dagger}(\omega)|0\rangle,
\end{equation}
simplifies things dramatically. Assuming that the transmission lines in the system and the output are in the vacuum state, and enforcing the condition for the input current, we find the following condition for the distribution $f(\omega)$,
\begin{equation*}
f(\omega) = \sqrt{\frac{4\pi}{\hbar\omega^{3}}}\frac{i I_{0}/2}{C_{g}\alpha^{*}\beta}\,\frac{\delta(\omega-\Omega)-\delta(\omega+\Omega)}{\sqrt{Z}s(\omega) - \sqrt{Z_{0}}(1+R_{0}(\omega))}.
\end{equation*}
Integrating this distribution, we are left with the following state of the source,
\begin{widetext}
\begin{equation}\label{state}
\begin{split}
|\psi\rangle =& \alpha \bigg\{ 1+ \sqrt{\frac{\pi\Omega}{\hbar Z_{0}}}\frac{I_{0}}{2 C_{g}\Omega^{2}|\alpha|^{2}}\bigg[\frac{1-i(C_{g}+C_{c}+C_{r})\Omega Z\theta - C_{g}\Omega Z_{0} (i + (C_{c}+C_{r})\Omega Z\theta)}{i+(C_{c}+C_{r})\Omega Z\theta + C_{r}\Omega Z_{1}(1-i C_{c}\Omega Z\theta)} \\
&- C_{r}\Omega Z_{1} \frac{i + (C_{g}+C_{c})\Omega Z\theta + C_{g}\Omega Z_{0} (1- i C_{c}\Omega Z\theta)}{i+(C_{c}+C_{r})\Omega Z\theta + C_{r}\Omega Z_{1}(1-i C_{c}\Omega Z\theta)}\bigg]a^{\dagger}(\Omega)\bigg\}|0\rangle,
\end{split}
\end{equation}
where parameter $\theta$ is defined in Eq.~\ref{theta}. For a quantized source modeled by a semi-infinite transmission line with the state presented above, we find the voltage in the system to be given described by
\begin{equation}\label{voltage}
\langle \dot{\phi}_{0} \rangle_{\psi} = I_{0}Z\theta \frac{\big(1+C_{r}^{2}\Omega^{2}Z_{1}(Z\theta+Z_{1})\big)\sin(\Omega t) - \Omega Z\theta (C_{c}+C_{r}+C_{c}C_{r}^{2}\Omega^{2}Z_{1}^{2}) \cos(\Omega t)}{1+ \Omega^{2}\big[ (C_{c}+C_{r})^{2}Z^{2}\theta^{2} + 2C_{r}^{2} Z\theta Z_{1} + C_{r}^{2}Z_{1}^{2} (1+C_{c}^{2}\Omega^{2}Z^{2}\theta^{2}) \big]},
\end{equation}
and the voltage in the output line by
\begin{equation}
\langle \dot{\phi}_{0}^{R} \rangle_{\psi} = I_{0}Z_{1}C_{r}\Omega Z\theta \frac{\big(C_{r}\Omega Z_{1} + (C_{c}+C_{r})\Omega Z\theta)\big)\sin(\Omega t) + (1 - C_{c}C_{r}\Omega^{2}Z\theta Z_{1}) \cos(\Omega t)}{1+ \Omega^{2}\big[ (C_{c}+C_{r})^{2}Z^{2}\theta^{2} + 2C_{r}^{2} Z\theta Z_{1} + C_{r}^{2}Z_{1}^{2} (1+C_{c}^{2}\Omega^{2}Z^{2}\theta^{2}) \big]}.
\end{equation}
\end{widetext}
Consequently, the outgoing current is defined as $I(t)\equiv\langle\dot{Q}^{R}_{0}\rangle_{\psi} = \langle \ddot{\phi}_{0}^{R} - \ddot{\phi}_{0}\rangle_{\psi}$, and it yields 
\begin{equation}
\langle \dot{Q}^{R}_{0} \rangle_{\psi} = \frac{\langle \dot{\phi}_{0}^{R} \rangle_{\psi}}{Z_{1}}.
\end{equation}
Note that we recover the results of Ref.~\cite{QHH} concerning the voltage response by setting $C_{r}=0$ and changing $Z_{0}\rightarrow Z\theta$. To complete this comparison, we compute the second moment of the voltage,
\begin{eqnarray}
\nonumber \langle\dot{\phi}_{0}^{2}\rangle &=& \frac{\hbar Z}{4\pi} \int_{0}^{\infty} d\omega \, \omega \Big[ |1+R(\omega)|^{2} + |s(\omega)|^{2}  \\
&+& |t(\omega)|^{2} \Big] + \frac{\hbar Z}{4\pi} \Omega |\alpha|^{2} |A(\Omega)|^{2} |s(\Omega)|^{2}, 
\end{eqnarray}
where we have defined $A(\Omega) = \beta \int d\omega \, f(\omega)$. In the same manner as for the single-channel Hodgkin-Huxley model, we regularize this integral by computing 
\begin{equation}
\Delta \equiv \langle\dot{\phi}_{0}^{2}\rangle_{\varphi_{1}} -\langle\dot{\phi}_{0}^{2}\rangle_{\varphi_{0}}
\end{equation}
with the choice of states $|\varphi_{1}\rangle = |\psi\rangle_{L}|\text{Thermal}\rangle|0\rangle_{R}$ and $|\varphi_{0}\rangle = |\psi\rangle_{L}|0\rangle|0\rangle_{R}$. This results in the quantity
\begin{equation}
\Delta = \frac{2 Z}{\pi \hbar \beta^{2}} ,
\end{equation}
a regularization of the zero-point energy in the circuit, up to second order in $\beta\hbar\omega\gg 1$, where $\beta=1/k_{B}T$. This leads us to believe that $\phi_{0}$ is a quantum variable.

\begin{figure*}[t]
\centering
{\includegraphics[width=1 \textwidth]{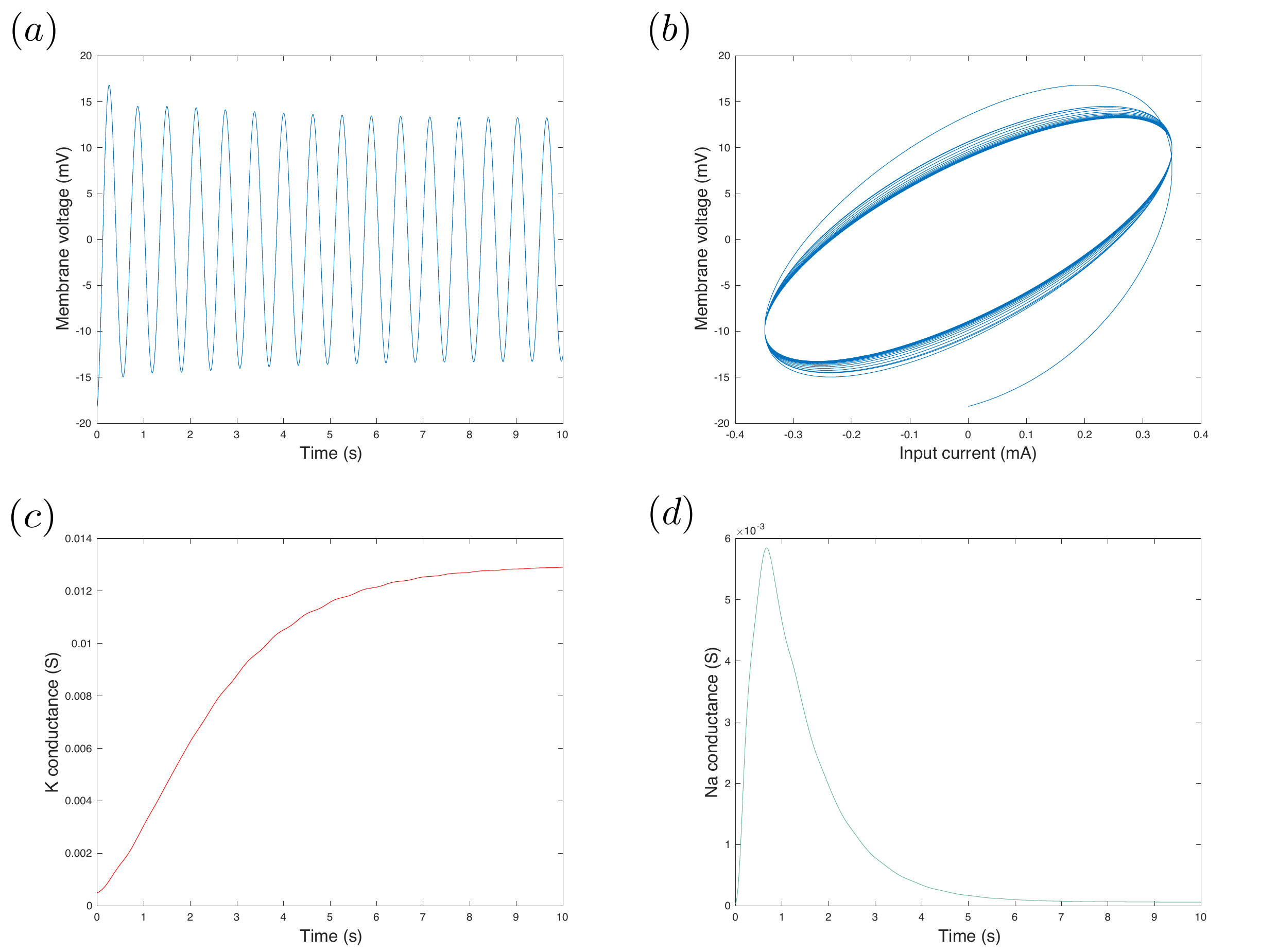}}
\caption{Quantized Hodgkin-Huxley model with potassium, sodium, and chloride ion channels, introducing a quantum source with state $|\psi\rangle$ and an output waveguide. (a) Membrane voltage over time. (b) Membrane voltage versus input current. (c)~Potassium channel conductance over time. (d) Sodium channel conductance over time. The membrane voltage is expressed in mV, the input current in mA, and the ion channel conductances in S.}
\label{fig5}
\end{figure*}

To observe memristive behavior of the system, we need to study the voltage-dependent dynamics of the impedances representing ion-channel conductances. These independent dynamics are described by
\begin{eqnarray}\label{Zupdate}
 \dot{Z}_{K} &=& -4 Z_{K}\Big(\frac{\alpha_{n}}{n} - (\alpha_{n}+\beta_{n})\Big),\\
\nonumber \dot{Z}_{Na} &=& -Z_{Na}\bigg[3\Big(\frac{\alpha_{m}}{m} - (\alpha_{m}+\beta_{m})\Big) \\
&+& \Big(\frac{\alpha_{h}}{h} - (\alpha_{h}+\beta_{h})\Big)\bigg],
\end{eqnarray}
where $\alpha_{i} = \alpha_{i}(V)$ and $\beta_{i}=\beta_{i}(V)$, for $i=n,m,h$.
Here, we have defined the impedances of the transmission lines modeling the ion channels as the inverse of their respective conductances, namely $1/g = Z$, given by
\begin{eqnarray*}
g_{K} = g_{K}^{max} n^{4} & \longrightarrow& Z_{K} = Z_{K}^{min} n^{-4},\\
g_{Na} = g_{Na}^{max} m^{3}h & \longrightarrow& Z_{Na} = Z_{Na}^{min} m^{-3}h^{-1}.
\end{eqnarray*}
These equations, together with Eq.~\ref{voltage}, allow for the computation of the voltage response in the system. Then, this solution can be used to consistently update the values of the impedances corresponding to the potassium and sodium channels. The validity of this approach relies on the assumption that the relaxation time of the set of LC oscillators which represents the instantaneous resistor is much shorter than the time scale associated to the change in the resistance, which is equivalent to an adiabatic approximation~\cite{Adiabatic}. We are essentially considering the impedances in the transmission lines to be constant in order to obtain the circuit voltage, which we afterwards use to update $Z_{K}$ and $Z_{Na}$ repeatedly. In this process, we assume that they change slowly in comparison with the circuit's relaxation time.

\section{Numerical Simulations}

\subsection{Quantized Hodgkin-Huxley model}
Here, we present the results for the voltage response in the system, as well as the potassium and sodium conductances, plotted over time (see Fig.~\ref{fig5}). These simulations are carried out through the update of the ion-channel conductances using the value of the voltage in the system, given by Eq.~\ref{Zupdate}, in an adiabatic approximation. The hysteresis loop that is displayed by the circuit voltage over the input current is also presented here. 

The circuit voltage is plotted over time in Fig.~\ref{fig5}~(a). We can observe similar dynamical behavior to what was found in the simulations of the single ion-channel Hodgkin-Huxley model, when studied in the quantum regime in Ref.~\cite{QHH}. The input current in the system, described by state $|\psi\rangle$ (in Eq.~\ref{state}) of the transmission line modeling the source, is $\langle \dot{Q}^{L}_{0}(t)\rangle_{\psi}\equiv I(t)=I_{0}\sin(\Omega t)$. 

\begin{figure}[]
\centering
{\includegraphics[width=0.45 \textwidth]{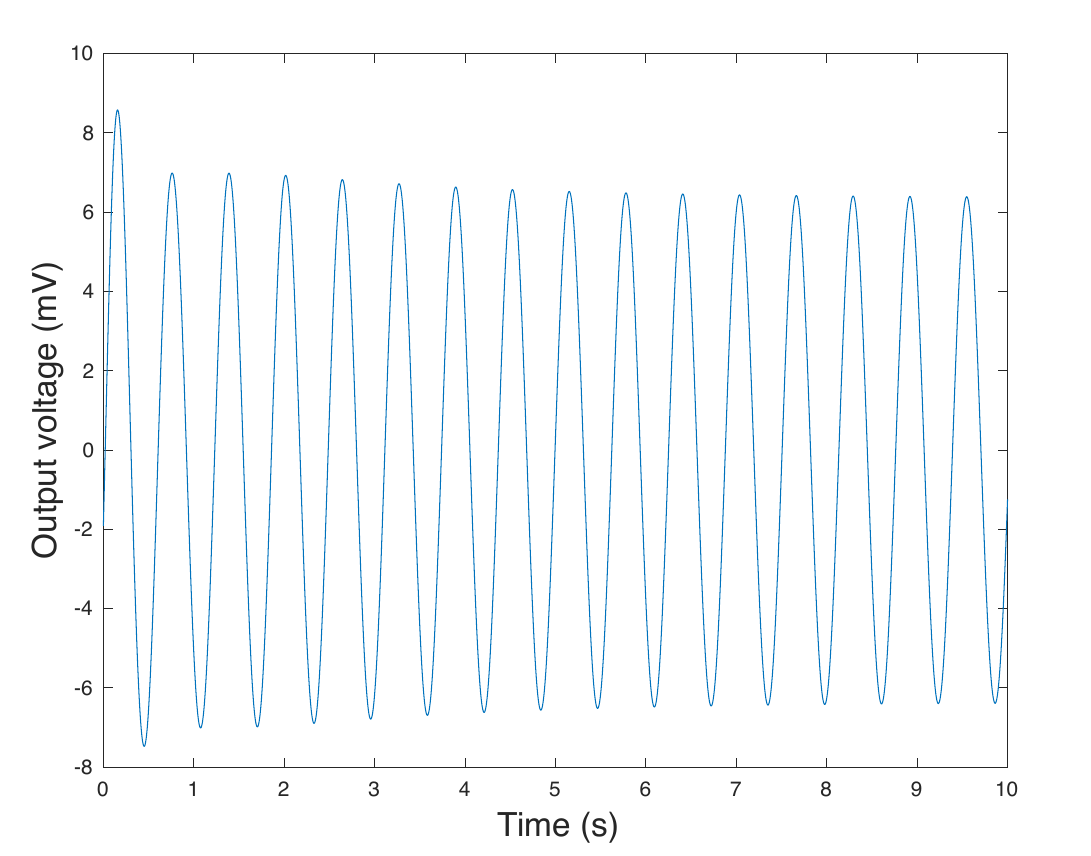}}
\caption{Voltage response over time in the output waveguide, for the quantized Hodgkin-Huxley model with potassium, sodium, and chloride ion channels, introducing a quantized source with state $|\psi\rangle$ and an output waveguide.}
\label{fig6}
\end{figure}

The I-V characteristic curve plotted in Fig.~\ref{fig5} (b) displays a hysteresis loop due to the periodic driving of the system, forming a limit cycle when the system saturates. It is remarkable that it is not pinched, opposite to the expected behavior of the classical memristor. The pinched hysteresis loop of the memristor appears when representing the voltage versus current, i.e. the response of the device versus the control variable. In the circuit that we are studying, apart from the memristors with different impedances describing the ion-channel conductances, the voltage response is affected by a capacitor. One of the effects of the presence of this capacitor is that, if it is initially charged, will displace the initial voltage on the circuit from zero (the neuron is no longer a passive circuit element). This change in behavior was also encountered when studying the quantization of a simplified version of the Hodgkin-Huxley model~\cite{QHH}, which comprises a single memristor to describe the potassium-ion channel.

\begin{figure}[t]
\centering
{\includegraphics[width=0.45 \textwidth]{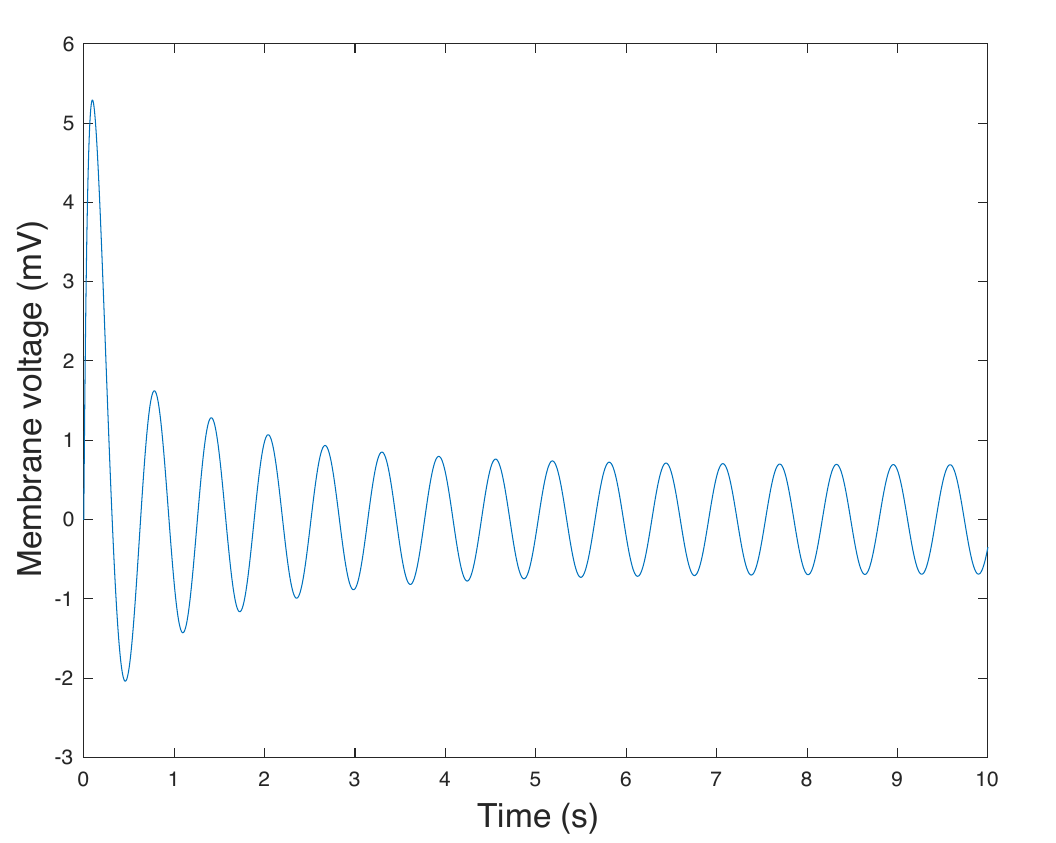}}
\caption{Membrane voltage over time in the quantized Hodgkin-Huxley model with potassium, sodium, and chloride ion channels, introducing a quantized source with state $|\psi\rangle$ and an output waveguide. This voltage response features spiking behavior, which was not found in the single-ion-channel model in an adiabatic regime.}
\label{fig7}
\end{figure}

In Fig.~\ref{fig5} (c) and (d), the potassium conductance features rising and adaptive behavior, and the sodium conductance shows a clear spike and relaxation. These behaviors are reproduced with great accuracy according to Ref.~\cite{HHM}. We can observe that, with the displacement of the membrane potential, the sodium activation channels switches on fast resulting in a spike in the conductance. Then, when potassium ions leave the axon in order to offset the increase of positive charge inside it, inactivation is switched on, and sodium ions gradually leave the cell. 

The voltage response in the output transmission line over time is represented in Fig.~\ref{fig6}. Its behavior features an initial spike and oscillations. However, this spike may simply be due to a decrease in amplitude.  

\subsection{Voltage spike}
Here, we present a choice of parameters in the circuit which can be tuned to reproduce the spike in the voltage, still in an adiabatic regime (see Fig.~\ref{fig7}). We consider the same state $|\psi\rangle$ of the quantum source, but now the current amplitude is $I_{0}=1$ mA, and the maximum values of the ion-channel conductances are $\bar{g}_{\text{K}} = 1.95$ S, $\bar{g}_{\text{Na}} = 0.69$ S, and $\bar{g}_{\text{Cl}} = 3\cdot 10^{-4}$ S. At $t=0$, the gate-opening probabilities for the potassium and sodium ion channels are $n=0.4$, $m=0.2$, and $h=0.6$. We have also set $\Omega=10$ Hz, $Z_{1}=50 \, \Omega$, and $C_{c}=C_{g}=C_{r}=10^{-6}$ F. By straying away from the biological model, we are able to observe a spike in the system's voltage response, which means that the action potential in neurons can be reproduced even in an adiabatic regime. 

\section{Conclusions \& Perspectives}
We have studied the quantization of the Hodgkin-Huxley model including potassium, sodium, and chloride ion channels, by means of introducing the concept of quantum memristor. The proposed setup consists of a quantum source coupled to the circuit and an output waveguide. This source is described by a semi-infinite transmission line whose state defines the input current in the system. The circuit comprises a capacitor, two quantum memristors and a resistor, modeling the potassium, sodium, and chloride ion channels, respectively. The quantum memristor is modeled by a semi-infinite transmission line with voltage-dependent impedance. 

The simulation of the voltage response in the circuit reveals similar dynamics to what was observed in Ref.~\cite{QHH} when studying the Hodgkin-Huxley model containing a single potassium channel. Concerning the potassium and sodium channel conductances, the results are in great accordance with the experiments in Ref.~\cite{HHM}. The sodium channel conductance shows a initial spike, followed by slow relaxation, consequence of the mechanism of the sodium channel, which consists of a fast activation gate followed by inactivation. During the latter, the activation gate of the potassium channel is switched on, and its conductance rises as a s-shaped curve. We also were able to reproduce spiking behavior of the membrane voltage in an adiabatic regime, but with a set of parameters which differs from the bio-inspired ones. This allows us to recover the spike of the action potential in the quantum regime.

In this work, we have proposed a state of the quantum source, and used a single-mode state to obtain the same harmonic input current as for the single-ion-channel Hodgkin-Huxley model with a classical AC source. When studying the second moment of the circuit voltage, we found a term related to the reflection of the modes in the circuit, which has a purely quantum-mechanical origin, related to the quantum fluctuations. This means that we are dealing with a quantum voltage. 

The introduction of multi-mode state could allow the study of the quantized Hodgkin-Huxley circuit in the presence of entanglement, and this is interesting for the construction of connected quantum neuron networks able to process quantum information. Thus, this setup sets an excellent starting point for advances in neuromorphic quantum computing with direct applications on hardware-based quantum machine learning.
\\
\acknowledgements
The authors acknowledge support from Spanish Government PGC2018-095113-B-I00 (MCIU/AEI/FEDER, UE) and Basque Government IT986-16. The authors also acknowledge support from the projects QMiCS (820505) and OpenSuperQ (820363) of the EU Flagship on Quantum Technologies, as well as the EU FET Open Grant Quromorphic. This work is supported by the U.S. Department of Energy, Office of Science, Office of Advanced Scientific Computing Research (ASCR) quantum algorithm teams program, under field work proposal number ERKJ333.  

\appendix
\section{State of the quantum source}
Here we derive the distribution $f(\omega)$ in the single-mode state of the source, according to the input current we want in the system, in order to compare with Ref.~\cite{QHH}. This state is
\begin{equation}
|\psi\rangle = \alpha|0\rangle + \beta\int d\omega f(\omega) a^{\dagger}(\omega)|0\rangle,
\end{equation}
with normalization given by
\begin{equation}
\langle\psi|\psi\rangle = |\alpha|^{2} + |\beta|^{2} \int d\omega |f(\omega)|^{2} = 1.
\end{equation}
We can consider $f(\omega)$ to be square-integrable function, $f(\omega)\in L^{2}$, leading to the normalization condition $|\alpha|^{2}+|\beta|^{2}=1$. We want to find the distribution $f(\omega)$ that satisfies
\begin{equation}
\langle \dot{Q}^{L}_{0}(t)\rangle_{\psi}=C_{g}\langle\ddot{\phi}^{L}_{0}(t)-\ddot{\phi}_{0}(t)\rangle_{\psi} = I_{0}\sin\Omega t.
\end{equation}
For that, we compute the action of these states on the modes describing the left transmission line,
\begin{equation}
\langle a^{L}_{\text{in}}(\omega)\rangle_{\psi} = \alpha^{*}\beta f(\omega).
\end{equation}
Considering the transmission lines that model the quantum memristor and the output waveguide to be in the vacuum state, we obtain
\begin{eqnarray}
\nonumber  && \langle\dot{Q}^{L}_{0}(t)\rangle = C_{g} \sqrt{\frac{\hbar}{4\pi}}\int_{0}^{\infty} \frac{d\omega}{\sqrt{\omega}} \omega^{2} \Big[ (s(\omega)\sqrt{Z} \\
&& -(1+R_{0}(\omega))\sqrt{Z_{0}})\alpha^{*}\beta f(\omega) e^{-i\omega t} + \text{H.c.} \Big]
\end{eqnarray}
Recall we imposed earlier $\langle\dot{Q}^{L}_{0}(t)\rangle_{\psi} \equiv I(t) = I_{0}\sin(\Omega t)$. This can be written as
\begin{equation*}
\langle\dot{Q}^{L}_{0}(t)\rangle_{\psi} = \int_{0}^{\infty} \frac{d\omega}{\sqrt{\omega}}\Big( \mathcal{I}(\omega)e^{-i\omega t} + \mathcal{I}^{*}(\omega)e^{i\omega t}\Big),
\end{equation*}
for $\mathcal{I}(\omega) = \frac{i I_{0}\sqrt{\omega}}{2}(\delta(\omega-\Omega) - \delta(\omega+\Omega))$. This leads to the identification
\begin{equation*}
f(\omega) = \sqrt{\frac{4\pi}{\hbar\omega^{3}}}\frac{i I_{0}/2}{C_{g}\alpha^{*}\beta}\,\frac{\delta(\omega-\Omega)-\delta(\omega+\Omega)}{\sqrt{Z}s(\omega) - \sqrt{Z_{0}}(1+R_{0}(\omega))}.
\end{equation*}

\section{Transmission and reflection coefficients}\label{appendix}
Here, we present the reflection and transmission coefficients between source, system, and output waveguides. Note that we find these relations for the transmission coefficients,
\begin{eqnarray}
\text{Source-System} \, &\longrightarrow& \, s(\omega) = \theta s_{0}(\omega),\\
\text{System-Output} \, &\longrightarrow& \, t(\omega) = \theta s_{1}(\omega),\\
\text{Source-Output} \, &\longrightarrow& \, t_{0}(\omega) = t_{1}(\omega).
\end{eqnarray}
The reflection coefficients are given by
\begin{widetext}
\begin{eqnarray}
\nonumber R_{0}(\omega) &=& \frac{1}{d(\omega)} \bigg\{ i + \omega\Big[(C_{g}+C_{c}+C_{r}) Z\theta - C_{g} Z_{0}(1-i(C_{c}+C_{r})\omega Z\theta) \Big] \\
&+& \omega C_{r} Z_{1}\big(1-i(C_{g}+C_{c})\omega Z\theta + C_{g}\omega Z_{0}(i+C_{c}\omega Z\theta)\big)\bigg\}, \\
\nonumber R(\omega) &=& \frac{1}{d(\omega)} \bigg\{ i + \omega\Big[(C_{g}+C_{c}+C_{r}) Z\theta - C_{g} Z_{0}(1+i(C_{c}+C_{r})\omega Z\theta) \Big] \\
&-& \omega C_{r} Z_{1}\big(1+i(C_{g}+C_{c})\omega Z\theta - C_{g}\omega Z_{0}(i-C_{c}\omega Z\theta)\big) \bigg\}, \\
\nonumber R_{1}(\omega) &=& \frac{1}{d(\omega)} \bigg\{ i + \omega\Big[(C_{g}+C_{c}+C_{r}) Z\theta + C_{g} Z_{0}(1-i(C_{c}+C_{r})\omega Z\theta)\Big] \\
&-& \omega C_{r} Z_{1}\big(1-i(C_{g}+C_{c})\omega Z\theta - C_{g}\omega Z_{0}(i+C_{c}\omega Z\theta)\big) \bigg\},
\end{eqnarray}
where we have defined
\begin{equation*}
d(\omega) = i + \omega\Big[(C_{g}+C_{c}+C_{r}) Z\theta + C_{g} Z_{0}(1-i(C_{c}+C_{r})\omega Z\theta) + C_{r} Z_{1}\big(1-i(C_{g}+C_{c})\omega Z\theta - C_{g}\omega Z_{0}(i+C_{c}\omega Z\theta)\big)\Big].
\end{equation*}
\end{widetext}
The transmission coefficients for the source-to-system case are given by
\begin{eqnarray}
s_{0}(\omega) &=& \frac{-2iC_{g}\omega\sqrt{Z_{0}Z}(i+C_{r}\omega Z_{1})}{d(\omega)},\\
s(\omega) &=& \frac{-2iC_{g}\theta\omega\sqrt{Z_{0}Z}(i+C_{r}\omega Z_{1})}{d(\omega)},
\end{eqnarray}
In the case of system-to-output, we have 
\begin{eqnarray}
s_{1}(\omega) &=& \frac{-2iC_{r}\omega\sqrt{Z Z_{1}}(i+C_{g}\omega Z_{0})}{d(\omega)},\\
t(\omega) &=& \frac{-2iC_{r}\theta\omega\sqrt{Z Z_{1}}(i+C_{g}\omega Z_{0})}{d(\omega)},
\end{eqnarray}
and for the source-to-output case,
\begin{eqnarray}
t_{0}(\omega) &=& \frac{-2 i C_{g}C_{r}\omega^{2}\theta Z\sqrt{Z_{0}Z_{1}}}{d(\omega)},\\
t_{1}(\omega) &=& \frac{-2iC_{g}C_{r}\omega^{2}\theta Z\sqrt{Z_{0}Z_{1}}}{d(\omega)}.
\end{eqnarray}

\end{document}